\documentclass[useAMS,usenatbib]{mn2e}

\newcommand{\dg}{$^\circ$}

\title[Geometric Calibration and Solar System Objects]{The Geometric Calibration of the {\it Planck} satellite using Solar System Objects}
\author[D. L. Harrison]{D. L. Harrison$^{1}$\thanks{E-mail: dlh@ast.cam.ac.uk (DLH)} and F. van Leeuwen$^{1}$ \\
$^{1}$ Institute of Astronomy, Madingley Road, Cambridge, CB3 0HA, UK\\}
\begin{document}

\date{}

\pagerange{\pageref{firstpage}--\pageref{lastpage}} \pubyear{2005}

\maketitle

\label{firstpage}

\begin{abstract}
The geometric calibration of the {\it Planck} satellite using the planetary transits is investigated, together with the reconstruction of any offsets from the nominal layout of the focal plane. The methods presented here may be applied to a single focal plane transit of a planet, to find the values of the geometric-calibration parameters at the epoch  of the transit or all the transits over the course of the mission. The pointing requirements are easily met, with the pointing reconstruction being dominated by the errors due to the star tracker.
\end{abstract}


\section{Introduction}
\label{intro}

{\it Planck} is a European Space Agency satellite designed to produce high-resolution temperature and polarisation maps of the CMB. It possesses detectors sensitive to a wide range of frequencies from 30 to 857 GHz, split between two instruments the HFI and LFI, the high and low frequency instruments, respectively.

{\it Planck} will be inserted into a Lissajous orbit around the second Lagrange point of the Earth-Sun system, spinning about its axis once per minute. The line of sight being almost perpendicular to the spin axis, hence the detectors will almost follow a great circle. The spin axis will nominally be repositioned every hour, and the roughly 60 or so circles corresponding to a single spin axis positioning may be binned together to form a ring. The nominal spin axis passes through the centre of the solar panels and is directed away from the sun, thus maintaining the rest of the satellite in a cone of shadow produced by the solar panels.  The scanning strategy is determined by the sequence of the nominal positions of the spin axis over the course of the mission. 

The geometric calibration is the process whereby all the line-of-sight positions of the detectors are recovered, at any time during the observations. The relationship between the pointing and time may depend on multiple parameters, discussed in full in \cite{leeuwen02}; here we discuss only those parameters which require calibrating with the science data. The remaining geometric-calibration parameters may be determined solely from the star tracker data.

The recovery of the geometric-calibration parameters has been investigated previously in \cite{harrison04} and \cite{harrison05}, which both use detections from point sources in the ring data to recover the geometric-calibration parameters during and post-mission, respectively.  In this paper we investigate the recovery of the geometric-calibration parameters using the solar system objects, specifically the planets. Additionally we show that it is possible to recover the focal plane layout, using the data corresponding to planetary transits.

Section~\ref{pointingRecon} contains an overview of the pointing reconstruction of the satellite, in terms of the geometric-calibration parameters. Methods to reconstruct these parameters and the focal plane layout from the detections due to the planets are presented in Section~\ref{methods_section}. The simulations used to assess the performance of these methods are described in Section~\ref{simul_section}. The results are presented in Section~\ref{results_section}, where it is shown that the geometric-calibration parameters and any offsets from the nominal focal plane layout may be recovered using the transits of the planets.

\section{Pointing Reconstruction}
\label{pointingRecon}

The pointing reconstruction of the satellite is achieved by the use of a star tracker. However, the exact relationship between the reference frame which contains the star tracker and the satellite reference frame which contains the focal plane is uncertain. This relationship must therefore be established using the science data in order to meet the pointing accuracy requirements as discussed in \cite{harrison04}. This may be achieved by the calibration of any offsets in the nominal values of the geometric-calibration parameters discussed below. A more detailed discussion of the reference systems and the attitude analysis may be found in \cite{leeuwen02} and \cite{harrison05}.

\begin{figure}
\begin{center}
\setlength{\unitlength}{1cm}
\begin{picture}(10,7.5)(0,0)
\put(-1.5,9.75){\includegraphics{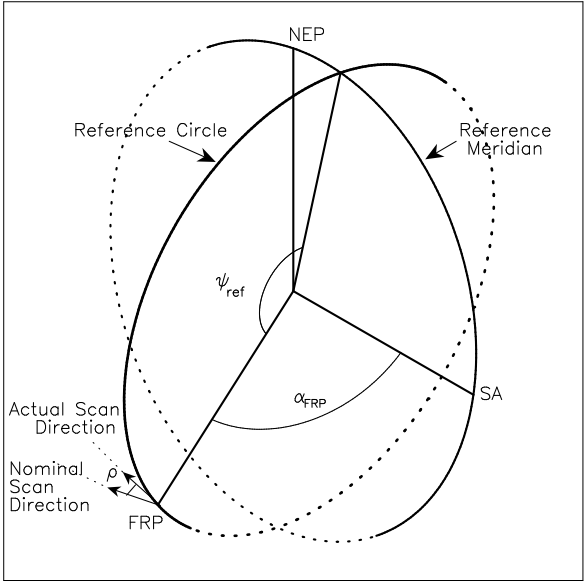}}
\end{picture}
\end{center}
\caption[]{The figure shows the geometric-calibration parameters which allow a description of the position of the field of view, FOV, with respect to the spin axis position, SA. The boresight angle, $\alpha_{FRP}$ is the angle between the FRP and the spin axis. The rotation of the focal plane around the FRP with respect to the nominal scan direction is given by the roll angle, $\rho$. The reference phase, $\psi_{ref}$, is the value of the initial phase at the point at which the FRP crosses the reference point, as defined by the intersection of the reference circle and reference meridian. Where the initial phase is measured from the first point observed on the reference circle, and the reference point is given by the intersection point of the reference circle and the great circle connecting the spin axis position a the NEP.}
\label{geo_fig}
\end{figure}

The geometric-calibration parameters are defined to allow the reconstruction of the lines-of-sight of the detectors with respect to the spin axis position, given the layout of the focal plane with respect to a reference point. This reference point may be defined as the centre of the focal plane, and is henceforth referred to as the fiducial reference point, FRP. The geometric-calibration parameters may then be defined by three angles, shown in Figure~\ref{geo_fig}. The boresight angle is defined by the angle between the spin axis and the FRP, and the roll angle by the rotation of the focal plane around the FRP. The reference phase defines the position along the reference circle, and is measured from a reference point defined by the intersection of the reference circle with the great circle connecting the spin axis position and the North Ecliptic Pole, NEP. A more detailed discussion of the geometric-calibration parameters may be found in \cite{harrison05}.

If the focal plane layout is uncertain then as well as the geometric-calibration parameters, offsets in the individual detector positions need to be recovered from the science data. These offsets may be expressed in terms of offsets in the nominal scan and cross-scan directions.

\section{Methods}
\label{methods_section}

The methods presented here are similar to those discussed in \cite{harrison05}, the main differences arise from the fact that the positions of the planets are known to high precision. As previously, the geometric-calibration parameters are assumed to be constant over the time frame of a ring and possess a slow linear variation over the course of the mission. The instantaneous offset in a geometric-calibration parameter, $\gamma$, may be defined on each ring by:
\begin{eqnarray}
\label{inst_offset_eqn}
\gamma \left(\Gamma_i\right) & = & \gamma_0 + \gamma_1 \cdot t_i \left( \Gamma_i\right) \nonumber \\
{\rm where,} \nonumber \\
t_i \left( \Gamma_i\right)  & = & \frac{\left( \Gamma_i-\Gamma_{max}/2 \right)}{\Gamma_{max}}
\end{eqnarray} where $\Gamma_i$ is the current ring number, $\Gamma_{max}$ is the final ring of the mission and the ring numbers start from zero. The systematic offset in the parameter, $\gamma_0$, is hence defined as the instantaneous offset of the parameter exactly half-way though the mission and the drift in the value of the parameter, $\gamma_1$, is defined as the total drift in the value of the parameter over the course of the mission.

First the offsets in the reference phase and roll angle are extracted, then the offsets in the boresight angle. The geometric-calibration parameters values are then iterated over until convergence is reached, only then is the focal plane layout is recovered.

\subsection{Reference Phase and Roll Angle}
\label{methods_refPhase}

Offsets in the reference phase and roll angle may be found from the differences between the observed and expected phases of the detections, where the phase difference for a detection is given by:
\begin{equation}
\label{phase_dif_eqn}
\Delta \psi = \psi_{o} -\psi_{e}
\end{equation} where $\psi_o$ is the observed phase of the detection and $\psi_e$ is the expected phase for the detection. The expected phase of a detection on the $i^{th}$ ring in the $d^{th}$ detector may be calculated using:
\begin{equation}
\label{expected_phase_eqn}
\psi_e = \frac{\sin(\beta_{p})-\cos(\alpha_i)\cos(\phi_i)}{\sin(\alpha_i) \sin(\phi_i)} +x_d(\rho_i)
\end{equation} where $\alpha_i$ is the angular separation between the position of the planet, ($\lambda_p$, $\beta_p$), and the spin axis position, ($\lambda_{SA}$, $\beta_{SA}$), at the occasion of the  $i^{th}$ ring, $\phi_i$ is the angle between the spin axis position and the NEP and $x_d$ is the scan position of the detector relative to the FRP.

The phase difference for a detection on the $i^{th}$ ring in the $d^{th}$ detector, $\Delta \psi_{id}$, may also be given by:
\begin{equation}
\label{lsq_eqn}
\Delta \psi_{id}  =  \psi_{ref_0} + \psi_{ref_1} \cdot t_i\left( \Gamma_i\right)  - y_d(\rho_i) \left( \rho_0 + \rho_1 \cdot t_i\left( \Gamma_i\right)  \right) 
\end{equation} where $y_d$ is the cross-scan position of the detector relative to the FRP at the time of the observation. The systematic offsets and drifts in the reference phase and roll angles may hence be solved for using a non-linear least squares analysis.

\subsection{The Boresight Angle}
\label{methods_openAng}

The ordinate, $\eta$, of a detection may be defined by the angular separation between the scan circle described by the detector and the position of the planet, and may be expressed as:
\begin{equation}
\label{ord_eqn}
\eta =\alpha_i - \alpha_d
\end{equation} where $\alpha_i$ is the angular separation between the planet and spin axis positions and $\alpha_d$ is the angle between the line-of-sight of the detector and the spin axis position and is given by:
\begin{equation}
\label{open_angle_eqn}
\alpha_d = \alpha_{FRP}\left( \Gamma_i\right)  + y_d(\rho_i)
\end{equation} where $\alpha_{FRP} \left( \Gamma_i\right) $ is the value of the boresight angle on the $i^{th}$ ring.

By accumulating detections from multiple rings, and hence fitting the cross-scan transit of the planet through the detector, the ordinate corresponding to the peak of the cross-scan transit may be found. This corresponds to the offset in the opening angle to this detector at the epoch of this planetary transit. The offset in the boresight is found for each planetary transit of the focal plane from the weighted mean of those offsets found for the detectors with the smallest beams, as in \cite{harrison05}. In practice this means the detectors which belong to the top four frequency channels. The systematic offset and drift in the boresight angle may be found by a straight line fit to the value found for each instantaneous offset from each focal plane transit.

\begin{equation}
\label{bore_eqn}
\alpha_{FRP} (t_p)  =  \alpha_{FRP_0} + \alpha_{FRP_1} \cdot t_p \left( \overline{\Gamma_p} \right)
\end{equation} where $\overline{\Gamma_p}$ is the mean ring number of the detections corresponding to those used in the evaluation of the boresight angle at the epoch of the planetary transit.

\subsection{Focal plane layout}
\label{methods_fp_layout}

After the geometric-calibration parameters have been successfully recovered, the focal plane layout with respect to the FRP may be extracted. An uncertainty in the focal length of the optics will produce offsets in the focal plane layout which scale as a function of the detector position relative to the FRP. 

The offset in the position of a detector with respect to the FRP may be expressed in terms of an offset in the nominal scan, $\delta x_{d,0}$, and in the nominal cross-scan, $\delta y_{d,0}$, directions. For each focal plane transit of a planet the offset in the detector position in the current scan and cross-scan directions are found. These may then be converted to offsets in the nominal scan and cross-scan direction using the value found for the roll angle at each epoch.
\begin{eqnarray}
\label{fp_eqn}
\delta x_{d,0} &=& \delta x_{d,\rho} \cos(\rho) - \delta y_{d, \rho} \sin(\rho) \nonumber \\
\delta y_{d,0} &=& \delta y_{d,\rho} \cos(\rho) + \delta x_{d, \rho} \sin(\rho)
\end{eqnarray}

The systematic offsets in the position of a detector may then be found from a weighted mean of the offsets found, in the nominal scan and cross-scan directions, for each focal plane transit.	

\section{Simulations}
\label{simul_section}

In order to assess the performance of the methods developed here in the reconstruction of the geometric-calibration parameters, it is necessary to simulate the planetary transits occurring during the {\it Planck} mission. This requires an ephemeris and the brightness temperatures of the planets in the {\it Planck} frequency bands as well as a scanning strategy and information on the beams of the {\it Planck} detectors. Throughout this paper we have made the simplifying assumption of Gaussian beams. These methods, however, may be easily extended to non-Gaussian beams, due to detector time constants or otherwise. Indeed, the beam profiles used are intimately connected to the geometric-calibration parameters, changes in the definition of the beam profiles will affect the geometric-calibration parameters. Hence, the geometric-calibration parameters are defined by the beam profiles used in their extraction.

\subsection{The scanning strategy}
\label{scanStrat}

While it is anticipated that this method will be applicable to any scanning strategy in which the circles corresponding to a single spin axis position may be binned together as a ring, only two potential scanning strategies for {\it Planck} were investigated here, a sinusoidal and a precessional scanning strategy. The sinusoidal scanning strategy may be described by:
\begin{eqnarray}
\label{Lsin_eqn}
\lambda_k & = & \lambda_0 + k \theta \nonumber \\
\beta_k & = & A \sin(n_s\lambda_k)
\end{eqnarray} and the precessional scanning strategy by:
\begin{eqnarray}
\label{Lprec_eqn}
\nu & = & (\lambda_0 + k\theta) \nonumber \\
\sin(\beta_k) & = & -\sin A \sin( n_p\nu) \nonumber \\
\cos(\phi) & = &\frac{\cos A}{\cos(\beta_k)} \nonumber \\
\lambda_k & = & \left \{ 
\begin{array}{ll} 
\nu+\phi &;\,\frac{\pi}{2} < n_p\nu < \frac{3\pi}{2}\\ 
\nu-\phi &;\,{\rm otherwise}\\ \end{array},
\right \}
\end{eqnarray} where, 
\begin{equation}
\lambda_0 = \lambda_\odot + \pi
\end{equation} and $\lambda_\odot$ is the position of the sun at the time the first ring, $k$ is the ring number, $\theta$ is the angular separation between subsequent spin axis positions, $n_{s,p}$ is the number of periods within $2\pi$, and $A$ is the amplitude. The values of these parameters used here are $\theta$=2.5\arcmin, $n_s$=2, $n_p=2.05$, and $A$=10\dg. This value of $\theta$, given a repointing once per hour, keeps the spin axis directed away from the sun.

\subsection{The Planets}
\label{planet_data}

An ephemeris code was written in order to evaluate the apparent positions and angular diameters of the planets as viewed from L2, the orbit of {\it Planck} about L2 is not included as this will not produce significant differences. The code uses the orbital elements available from JPL ({\it http://ssd.jpl.nasa.gov/elem\_planets.html}).

Planetary brightness temperatures from  \cite{dePater89}, \cite{griffin93}, \cite{griffin86}, \cite{hildebrand85}, \cite{loewenstein77}, \cite{muhleman91}, \cite{orton86}, \cite{rather74}, \cite{ulich76}, \cite{ulich74}, \cite{ulich73} and  \cite{whitcomb79} at far-infrared and sub-mm frequencies were fitted using a third order polynomial fit, as in \cite{griffin93}, except in the case of Mars in which a straight line fit is sufficient. The brightness temperatures thus found  for each planet, in each of the {\it Planck} frequency channels are shown in Table~\ref{brightness_temp_table}. These brightness temperatures may be used to calculate the expected flux of each detection due to each planet, given the angular diameter of the planet at the time of the observation as provided by the ephemeris data.

\begin{table}
\caption{The brightness temperatures at each {\it Planck} frequency, used in the simulations to generate the observed fluxes for the detections due to the planets.}
\label{brightness_temp_table}
\begin{tabular}{|rrrrrr|}
\hline
 Freq & Mars & Jupiter & Saturn & Uranus & Neptune \\
 (GHz) & (K) & (K) & (K) & (K) & (K) \\
\hline
30 & 211.7 & 172.4 & 143.9 & 139.7 & 135.8 \\
44 & 212.0 & 172.8 & 144.7 & 135.8 & 131.8 \\
70 & 212.5 & 173.2 & 146.0 & 129.0 & 124.8 \\
100 & 213.1 & 173.4 & 146.9 & 121.8 & 117.5 \\
143 & 213.9 & 173.3 & 147.4 & 112.8 & 108.3 \\ 
217 & 215.4 & 172.0 & 146.2 & 100.3 &  95.6 \\ 
353 & 218.0 & 167.0 & 139.0 & 85.5 & 80.6 \\ 
545 & 221.8 & 157.7 & 124.0 & 75.9 & 71.7 \\ 
857 & 228.0 & 149.6 & 112.7 & 62.6 & 62.2 \\ 
\hline
\end{tabular}
\end{table}

\subsection{Generating simulated data}
\label{simGen}

As in \cite{harrison05} we directly simulate the list of detections, as delivered by the Level 2 DPC from their analysis of the time ordered data, TOD. This list of detections includes the position in phase and the amplitude observed for the transit, together with their respective errors. Additionally, the list includes the number of the detector which made the observation and the ring number on which it occurred.  

The simulations assume that the spin axis is repositioned every hour and that the nutation effects are small so that the individual scans may indeed be combined into rings. For every spin axis position, the instantaneous values of the geometric-calibration parameters are assessed and used to reconstruct the lines-of-sight of the focal plane. If a planet is located nearby, the amplitude with which it is observed is evaluated and if above a threshold signal-to-noise ratio it may be included in the simulated list of detections. Errors on the amplitude of the detections are generated assuming the goal values of the noise in the ring, as in \cite{harrison05}, and white noise.

 The magnitude of the error in the phase of the detection, $\sigma_{\psi}$, is assessed using the empirical relationship found in \cite{harrison05}. However, this relationship must be modified to include the error in the velocity-phase relation, from the star tracker, which dominates for the highest signal-to-noise detections. The value of  $\sigma_{\psi}$ is now given by:
\begin{equation}
\label{phase_err_eqn}
\sigma_{\psi}^2= \left (\frac{1.7}{SNR}\right)^2 \sigma_b^2 + \sigma_{\nu}^2 
\end{equation} where SNR is the signal-to-noise ratio of the detection, $\sigma_b$ is the beam sigma of the detector and $\sigma_{\nu}$ is the error in the velocity-phase relation, recovered from the star tracker data.

\subsubsection{Saturation of detectors}
\label{saturation}

Jupiter may saturate the detectors; the effects of the potential saturation of the transits are considered here. Any saturation effects will only occur in extremely high signal-to-noise transits. The position of the peak of these transits in the scan direction is dominated by the error in the velocity-phase relation so the exclusion of the saturated measurements around the peak of the transit does not affect the accuracy at which the position of the peak of the transit may be found.

As the planets are in the ecliptic plane the relatively large angular separations between successive rings in the cross-scan direction has the effect that only a small proportion of the observations in a cross-scan transit are saturated. If the saturated measurements are discarded then the position of the peak of the cross-scan transit may be successfully recovered. This again does not have a significant effect on the accuracy of the position of the peak of the transit.

These effects were investigated using single scan and cross-scan transits; as the effects are not significant for any reasonable saturation limits they have not been included in the full simulations.

\section{Results}
\label{results_section}

\begin{figure*}
\begin{center}
\setlength{\unitlength}{1cm}
\begin{picture}(17.5,13)(0,0)

\put(9.5,6){\includegraphics{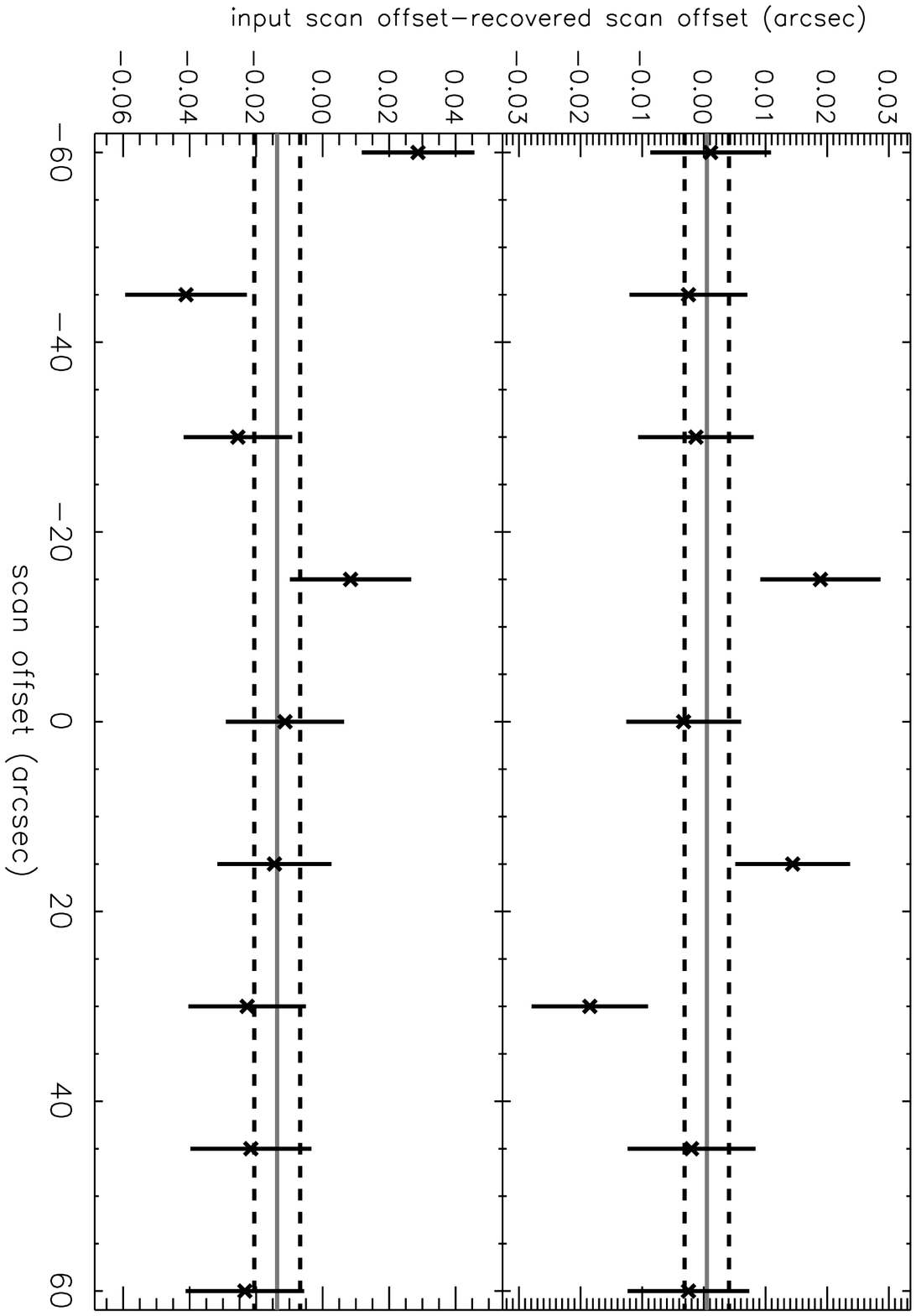}}

\put(18.5,6){\includegraphics{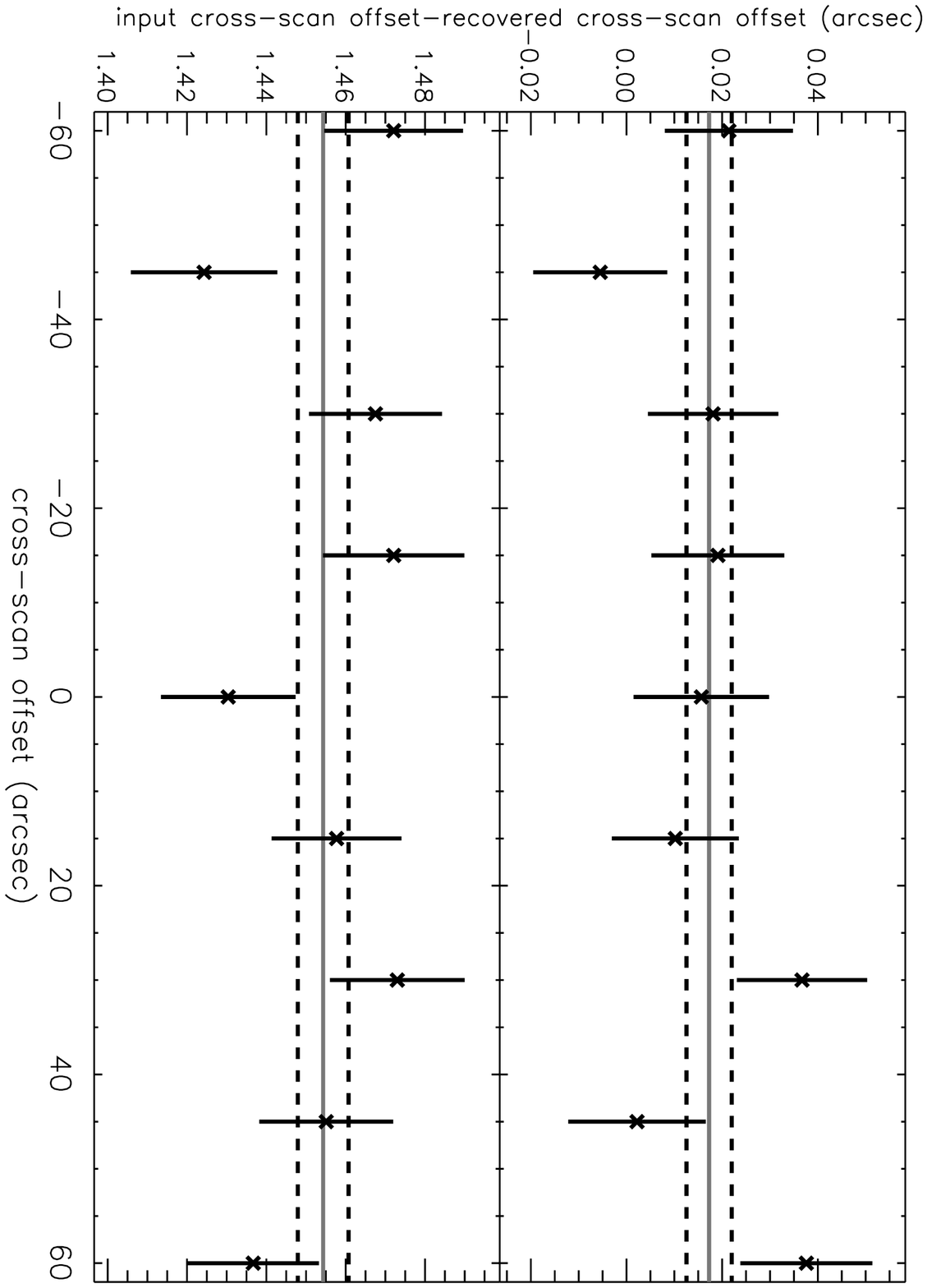}}

\put(9.5,-0.5){\includegraphics{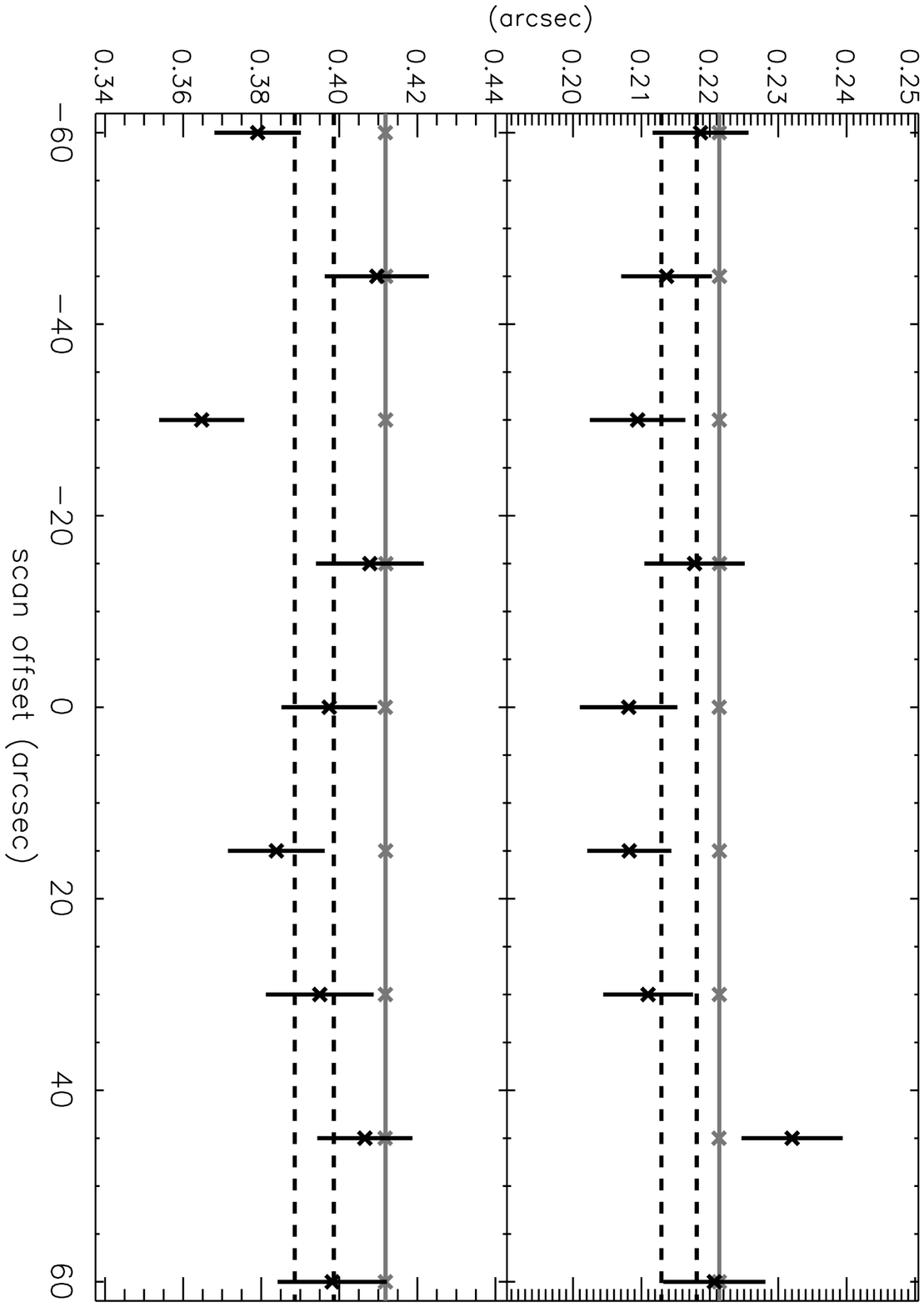}}

\put(18.5,-0.5){\includegraphics{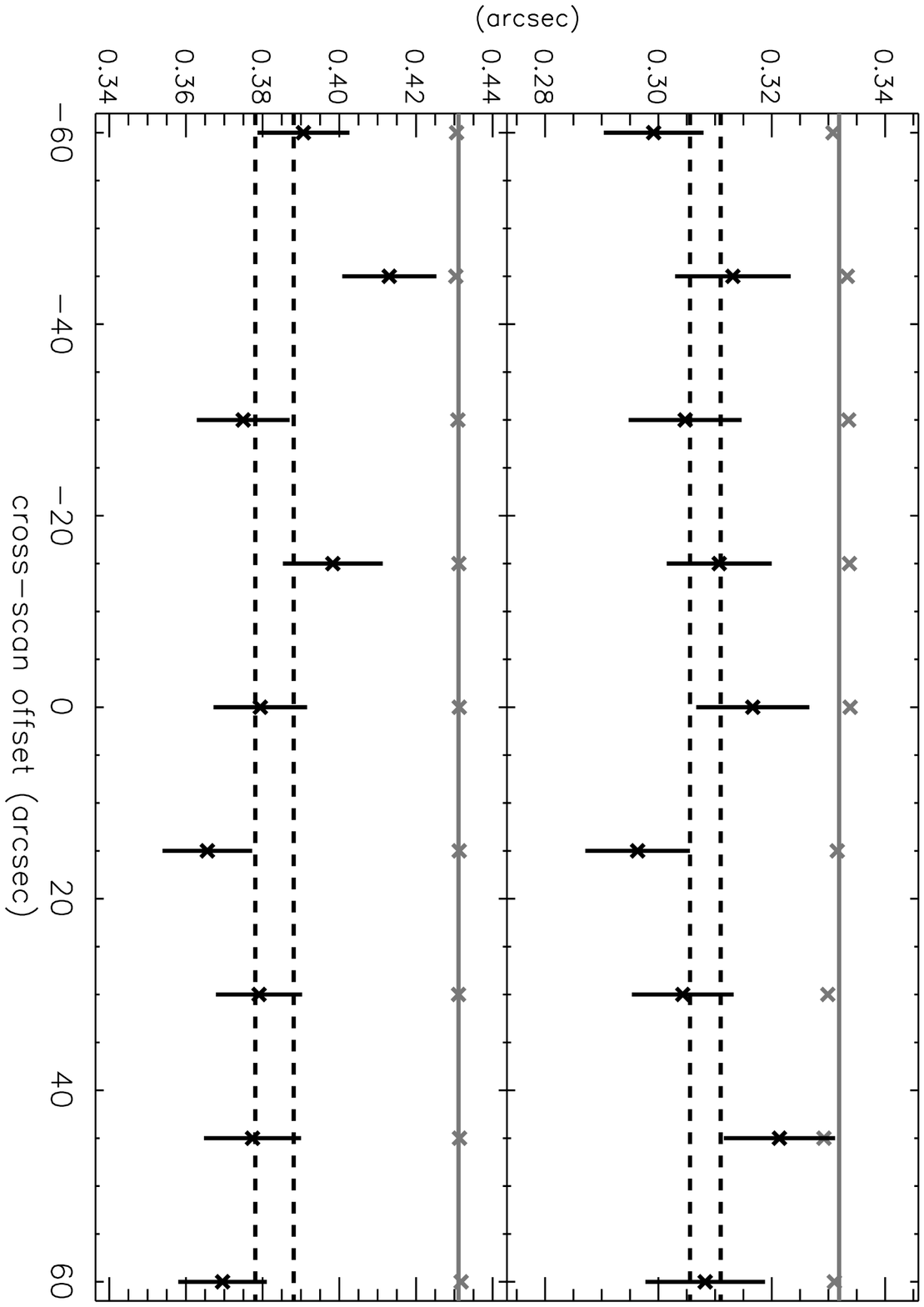}}

\end{picture}
\end{center}

\caption[]{These plots relate to the recovery of the line-of-sight, LOS, of a detector in the nominal scan and cross-scan directions, with the top panel in each plot containing the results for a 217~GHz, HFI detector and the bottom panel a 30~GHz LFI detector. 
Top left: the difference between the input and recovered value of the offset in the LOS of the detector in the nominal scan direction is plotted against the input value. The grey line is the global mean of these differences and the dashed lines enclose the region with 1$\sigma$ of this global mean.
Top right: the difference between the input and recovered value of the offset in the LOS, in the cross-scan direction, is plotted against the input value. The grey line is the global mean of these differences and the dashed lines enclose the region with 1$\sigma$ of this global mean. The recovery of the offset in the cross-scan LOS, may hence be seen to be biased by 0.02\arcsec\, in the case of the 217~GHz detector and 1.45\arcsec\, in the case of the 30~GHz detector.
Bottom left: shows the dispersion in the mean recovered value of the offset in the nominal scan direction, black crosses, and the mean calculated error of the error in the recovered offset in the nominal scan direction, grey crosses, against the actual value of the scan offset. The dashed black lines enclose the 1$\sigma$ region about the mean dispersion, and the grey line is the mean value of the calculated error.
Bottom right: shows the dispersion in the mean recovered value of the offset in the nominal cross-scan direction, black crosses, and the mean calculated error of the error in the recovered offset in the nominal cross-scan direction, grey crosses, against the actual value of the cross-scan offset. The dashed black lines enclose the 1$\sigma$ region about the mean dispersion, and the grey line is the mean value of the calculated error. The calculated value of the error in the cross-scan LOS offset is seen to be an overestimate of the actual error in the recovered values.
}
\label{offsets_fig}
\end{figure*}

\begin{table}
\caption{The errors in the offsets in the geometric-calibration parameters recovered at each of these planetary transits of the focal plane.}
\label{singlePlanet_table}
\begin{tabular}{|lrrrrr|}
\hline
Planet:& Mars & Jupiter & Saturn & Uranus & Neptune \\
Date :&03/08 & 04/08 & 05/08 & 06/08 & 05/08 \\
 &  (\arcsec) & (\arcsec) &  (\arcsec) & (\arcsec) & (\arcsec) \\
\hline
$\sigma_{\psi_{ref}}$   & 0.06 & 0.06 & 0.07 & 0.16 & 0.28 \\
$\sigma_{\alpha_{FRP}}$ & 0.17 & 0.23 & 0.23 & 0.38 & 0.62 \\
$\sigma_{\rho}$       &  3.71 & 1.98 & 3.51 & 12.6 & 25.2 \\
\hline
\end{tabular}
\end{table}

The simulated list of detections may then be analysed using the methods presented in Section~\ref{methods_section}. Unless otherwise stated 500 noise realisations are performed for each simulation. The magnitude of the errors used are as follows, the velocity-phase relation, $\sigma_{\nu}$, and errors in the spin axis position coordinates,  $\sigma_{\beta_{SA}}$ and $\sigma_{\lambda_{SA}}$, are all assumed to be 1\arcsec. The nutation amplitude used in the simulations was 1\arcsec and the minimum signal-to-noise ratio of a detection used in the analysis is 30.

Given the large number of high signal-to-noise ratio detections as a result of a planetary transit of the focal plane, and the high accuracy to which the positions of the planet are known it is possible to solve for the instantaneous values of the geometric-calibration parameters at the epoch of the focal plane transit. Table~\ref{singlePlanet_table} shows the errors in the recovered values of the geometric-calibration parameters at the epoch of a focal plane transit of each of the planets visible by {\it Planck}. Table~\ref{singlePlanet_table} also shows that if there are only systematic offsets in the geometric-calibration parameters then a single transit of any planet will reach the required pointing accuracy. The transits of Uranus and Neptune do not illuminate all the focal plane, which is the reason behind the large increase in the error in the roll angle. Due to the relative motions of {\it Planck} and Mars there are large variations in the time between successive transits of Mars, the first and potentially last opportunity to observe Mars transit the focal plane is March 2008. After which the next time Mars will be in the right relative location, will be in October 2009, after the end of the mission.

\begin{table}
\caption{Comparing the errors in the recovered values of the geometric-calibration parameters in the cases of the sinusoidal and precessional scanning strategies. In both cases only detections with signal-to-noise ratios of 30 or greater were used in the analysis.}
\label{meanCalERRS_precVsin_table}
\begin{tabular}{|lrr|}
\hline
Scanning Strategy: & sinusoidal & precessional \\
 &  (\arcsec) & (\arcsec) \\
\hline
$\sigma_{\psi_{ref_0}}$   & 0.03 & 0.03 \\
$\sigma_{\psi_{ref_1}}$   & 0.13 & 0.13 \\
$\sigma_{\alpha_{FRP_0}}$ & 0.11 & 0.10 \\
$\sigma_{\alpha_{FRP_1}}$ & 0.49 & 0.47 \\
$\sigma_{\rho_{0}}$       & 1.20 & 1.19 \\
$\sigma_{\rho_{1}}$       & 4.82 & 5.27 \\
\hline
\end{tabular}
\end{table}

Table~\ref{meanCalERRS_precVsin_table} shows the errors in the recovered values of the systematic offsets and the drifts in the geometric-calibration parameters using all the planetary focal plane transits over the course of the mission. Where the scanning strategy is assumed to start in July 2008, meaning no transit of Mars is included. Using an earlier start date for the scanning strategy, in order to include the March 2008 transit of Mars, does not however significantly improve the values displayed. Table~\ref{meanCalERRS_precVsin_table} also shows that the scanning strategy employed has virtually no impact on the ability of these methods to recover the geometric-calibration parameters, with no significant differences found between the errors in the recovered values in the cases of the sinusoidal and precessional scanning strategies.

\begin{table}
\caption{The maximum error in the line-of-sight of an HFI and LFI detector, due to the errors in the recovery of the geometric-calibration parameters shown in Table~\ref{meanCalERRS_precVsin_table}}
\label{geoCal_pointingERR_table}
\begin{tabular}{lrr}
\hline
Scanning Strategy: & sinusoidal & precessional \\
 & (\arcsec) & (\arcsec) \\
\hline
HFI (545 GHz) & 0.29 & 0.28 \\
LFI (44 GHz) & 0.36 & 0.36 \\
\hline
\end{tabular}
\end{table}

Table~\ref{geoCal_pointingERR_table} shows the effect of the errors in the geometric-calibration parameters in Table~\ref{meanCalERRS_precVsin_table} on the reconstruction on the line-of-sight of the {\it Planck} detectors. The maximum error in the line-of-sight will occur at the beginning and end of the mission due to the higher sensitivity to errors in the drift in the geometric-calibration parameters. The location of the detector in relation to the FRP will also affect the error in its line-of-sight due to the uncertainties in the roll angle. Table~\ref{geoCal_pointingERR_table} shows the largest errors found in the line-of-sight of the HFI and LFI detectors, corresponding to the values in Table~\ref{meanCalERRS_precVsin_table}.

\begin{figure}
\begin{center}
\setlength{\unitlength}{1cm}
\begin{picture}(8.5,6.5)(0,0)
\put(9.5,0){\includegraphics{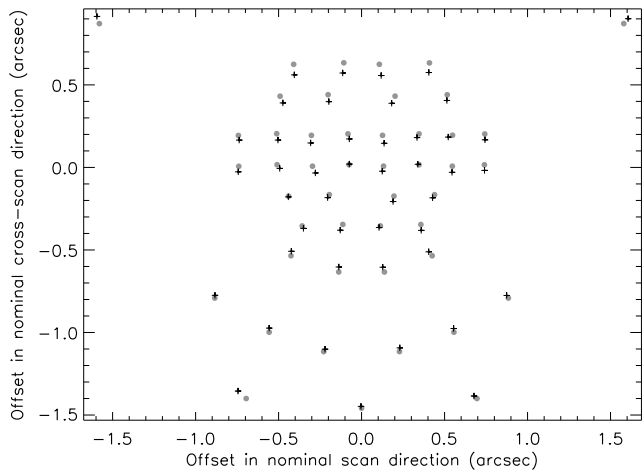}}
\end{picture}
\end{center}
\caption[]{The filled grey circles are the offsets in the detector positions. These offsets are 0.01\% of the detectors position wrt to the FRP. The black crosses are the mean recovered positions found for the detectors with the 1$\sigma$ errors in the mean recovered value. There remain some small biases in the recovery of the detectors, especially in the nominal cross-scan direction of the order of a few hundredths of an arcsecond.}
\label{fp_bias_fig}
\end{figure}

Figure~\ref{offsets_fig} demonstrates the ability of the methods presented here to recover the offsets in the line-of-sight of a detector. Offsets in the line-of-sight position of a detector with respect to the FRP are varied to assess the performance of the method. In each case the geometric-calibration parameters are also recovered, and are consistent with the zero offsets input to the simulations. The top panel in each plot, in Figure~\ref{offsets_fig}, corresponds to a 217~GHz HFI detector whereas the bottom panel corresponds to a 30~GHz LFI detector. The top left plot in Figure~\ref{offsets_fig} shows the difference between the input and recovered value of a line-of-sight offset in the nominal scan direction against the value input to the simulations, whereas the top right plot shows the difference between the input and recovered value of a line-of-sight offset in the nominal cross-scan direction against the value input to the simulations. In both plots the grey line shows the global mean of the differences and the dashed lines enclose the 1$\sigma$ region around the global mean. It may be seen in the top right plot in Figure~\ref{offsets_fig} that the recovered value of the offset in the nominal cross-scan direction for each of the detectors is biased. This bias was found to depend on the number of planetary focal plane transits used in the analysis and the beam size of the detector, with the largest biases seen in detectors with the largest beams. It is this bias in the recovered cross-scan position which necessitates the limitation on the evaluation of the boresight angle to the top four frequency channels, which have the smallest beams. In the top right plot, the 217~GHz detector, FWHM of 5\arcmin, is seen to be biased at the level of 0.02\arcsec\, whereas the 30~GHz detector, FWHM of 34.44\arcmin, is found to be biased at the level of 1.45\arcsec. Given that the magnitude of the bias is constant given the beam size and the planetary transits observed, it may be evaluated and removed if necessary, though at these magnitudes this is not required. The bottom left plot in Figure~\ref{offsets_fig} shows the errors in the recovered values of the offset in the nominal scan direction against the input offsets. This figure shows both the mean calculated values of the errors, grey crosses, and the dispersion in the recovered value of the offset, black crosses. The dashed lines enclose the 1$\sigma$ region around the global mean of the dispersion, and the grey line shows the global mean of the mean calculated error.  The bottom right plot in Figure~\ref{offsets_fig} shows the errors in the recovered values of the offset in the nominal cross-scan direction, again both the dispersion in the recovered values and the mean calculated errors are shown. Here, however, the mean calculated errors are seen to be overestimated with respect to the dispersion in the recovered values. In both these figures there is seen to be no dependence on the input value of the offset and the error in the recovered value.

Since the bias in the recovery of the nominal cross-scan position of a detector is constant given the planetary transits observed, the bias may be removed. This is shown in Figure~\ref{fp_bias_fig} were the input offsets, filled grey circles, in the detector positions are determined as a function of their distance from the FRP. These offsets are successfully recovered and any remaining biases are of the level of a few hundredths of an arcsecond. The black crosses show the recovered position of the detectors and the 1$\sigma$ error in that position.

\begin{table}
\caption{The maximum error in the line-of-sight of an HFI and LFI detector, due to the errors in the recovery of the geometric-calibration parameters shown in Table~\ref{meanCalERRS_precVsin_table} including the errors in the positions of the detectors with respect to the FRP.}
\label{los_pointingERR_table}
\begin{tabular}{lrrrr}
\hline
 & \multicolumn{2}{c}{sinusoidal} & \multicolumn{2}{c}{precessional} \\
 & cross-scan & scan & cross-scan & scan \\
 & (\arcsec) & (\arcsec) & (\arcsec) & (\arcsec) \\
\hline
HFI (217 GHz) & 0.34 & 0.22 & 0.32 & 0.21 \\
LFI (30 GHz) & 0.43 & 0.41 & 0.44 & 0.44 \\
\hline
\end{tabular}
\end{table}
The total error in the line-of-sight of a detector includes the error due to its recovered position relative to the FRP. Table~\ref{los_pointingERR_table} shows the largest errors found in the line-of-sight of the HFI and LFI detectors, once the errors in the positions of the detectors relative to the FRP are included.

\section{Discussion}
\label{discussion_section}

The geometric-calibration parameters and focal plane layout may be recovered with exquiste accuracy from the planetary transits of the focal plane. However, the transits of the focal plane by the planets are far from uniformly distributed in time. Depending on the start date of the mission there may a period of $\sim$ 5 months without a planetary transist and it is possible that this period may occur at the end of the mission. If the geometric-calibration parameters vary with time then even if this is only a linear variation, it may be necessary to confirm this variation using the bright extragalactic point sources, as in \cite{harrison05}. An independent measurement of the geometric-calibration parameters may also be highly desirable in the case of an alteration to the scanning strategy,  in order to improve the mapping of the beams of the detectors, when observing the planets

\section*{Acknowledgments}
This work was supported by PPARC at the Cambridge Planck Analysis Centre.


\label{lastpage}

\begin{thebibliography}{99}

\bibitem[\protect\citeauthoryear{de Pater et. al.}{1989}]{dePater89} de Pater I., Ulich B.L., Kreysa E., Chini R., 1989, Icarus, 79, 190

\bibitem[\protect\citeauthoryear{Griffin \& Orton}{1993}]{griffin93} Griffin M.J., and Orton G.S., 1993, Icarus, 105, 537

\bibitem[\protect\citeauthoryear{Griffin et. al.}{1986}]{griffin86} Griffin M.J., Ade P.A.R., Orton G.S., Robson E.I., Gear W.K., Nolt I.G., Radostitz J.V., 1986, Icarus, 65, 244

\bibitem[\protect\citeauthoryear{Harrison \& van Leeuwen}{2005}]{harrison05} Harrison D.L., van Leeuwen F. 2005, MNRAS, submitted

\bibitem[\protect\citeauthoryear{Harrison, van Leeuwen \& Rowan-Robinson}{2004}]{harrison04} Harrison D.L., van Leeuwen F., Rowan-Robinson M., 2004, MNRAS, 348, 1241

\bibitem[\protect\citeauthoryear{Hildebrand et. al.}{1985}]{hildebrand85} Hildebrand R.H., Loewenstein R.F., Harper D.A., Orton G.S., Keene J., Whitcomb S.E., 1985, Icarus, 64, 64

\bibitem[\protect\citeauthoryear{Loewenstein et. al.}{1977}]{loewenstein77} Loewenstein et. al., 1977, Icarus, 31, 315 

\bibitem[\protect\citeauthoryear{Muhleman \& Berge}{1991}]{muhleman91} Muhleman D.O, Berge G.L., 1991, Icarus, 92, 263

\bibitem[\protect\citeauthoryear{Orton et. al.}{1986}]{orton86} Orton G.S., Griffin M.J., Ade P.A.R., Nolt I.G., Radostitz J.V., 1986, Icarus, 67, 289

\bibitem[\protect\citeauthoryear{Rather, Ulich \& Ade}{1974}]{rather74} Rather J.D.G., Ulich B.L., Ade P.A.R., 1974, 22, 448

\bibitem[\protect\citeauthoryear{Ulich}{1974}]{ulich74} Ulich B.L., 1974, Icarus, 21, 254

\bibitem[\protect\citeauthoryear{Ulich \& Conklin}{1976}]{ulich76} Ulich, B.L., Conklin, E.K., 1976, Icarus, 27, 183

\bibitem[\protect\citeauthoryear{Ulich, Cogdell \& Davis}{1973}]{ulich73} Ulich B.L., Cogdell J.R., Davis J.H., 1973, Icarus, 19, 59

\bibitem[\protect\citeauthoryear{van Leeuwen et al.}{2002}]{leeuwen02} van Leeuwen, F., et al., 2002, MNRAS, 331, 975 

\bibitem[\protect\citeauthoryear{Whitcomb et. al.}{1979}]{whitcomb79} Whitcomb S.E., Hildebrand R.H., Keene J., Stiening R.F., Harper D.A.,1979, Icarus, 38, 75

\end{thebibliography}
\end{document}